\documentclass[a4paper,11pt]{article}
\usepackage{latexsym}
\usepackage{subfigure}
\usepackage{amsmath}
\usepackage{nicefrac}
\usepackage{JASA_manu}

\usepackage{tabularx}
\usepackage{colortbl}
\usepackage{hhline}
\usepackage{rotfloat}
\usepackage{subfig}
\usepackage{soul}
\usepackage{bbm}
\usepackage{multirow}
\usepackage{color}
\usepackage{framed}
\usepackage{comment}
\usepackage{apacite}
\definecolor{shadecolor}{gray}{0.875}
\includecomment{comment} 

\specialcomment{comment}{\begin{shaded}}{\end{shaded}}

\usepackage{natbib}
\usepackage{amssymb}
\title{A  SEIR model with time-varying coefficients for analysing the SARS-CoV-2 epidemic}
\author{Paolo Girardi\thanks{
		Address for correspondence: Paolo Girardi,
		Department of Developmental and Social Psychology - University of Padova,
		Via Venezia 8, 35131 Padova - Italy. 		E-mail: \texttt{paolo.girardi@unipd.it} }$^{\mbox{ }1,2}$
		and
	Carlo Gaetan$^3$ \\
	\noindent
	$^1$Department of Developmental and Social Psychology, 	University of Padova, Italy\\
	$^2$Department of Statistical Sciences, 	University of Padova, Italy\\
		$^3$Department of Environmental Sciences,  Informatics and Statistics, \\
	Ca' Foscari University of Venice, Italy
}

\begin{document}
\maketitle
\textbf{Abstract}\\
In this study, we propose a time-dependent Susceptible-Exposed-Infected-Recovered (SEIR)  model for the analysis of the SARS-CoV-2 epidemic outbreak in three different countries, the United States of America, Italy and Iceland using public data inherent the numbers of the epidemic wave. Since several types and grades of actions were adopted by the governments, including travel restrictions, social distancing, or limitation of movement, we want to investigate how these measures can affect the epidemic curve of the infectious population. The parameters of interest for the SEIR model were estimated employing a composite likelihood approach. Moreover, standard errors have been corrected for  temporal dependence. The adoption of restrictive measures results in flatten epidemic curves, and the future evolution indicated a decrease in the number of cases.\\

\textbf{Short summary}\\
A time-dependent SEIR model for the analysis of the SARS-CoV-2 epidemic outbreak using public data. Statistical methods allow understanding contagion dynamics under uncertain conditions.
\clearpage

\section{Introduction}\label{sec:introduction}
The use of epidemic models permits to simulate disease transmission dynamics to detect emerging outbreaks and to assess public health interventions \citep{unkel2012,boily2007}. 
With the scope to describe the dynamics of epidemics, standard methods, such as the SIR model \citep{anderson1992}, divide the population  into portions of subjects on the basis of their relation concerning the epidemic vector. Here the focus is on the dynamic such as the depletion of the susceptible portion to the infected one or the possible evolution of the rate of immunization.

However, the standard SIR model, and other extensions as the SEIR model, do not take into account the time-varying nature of epidemics and several attempts were made to overcome this limitation \citep{Dureau:2013,boatto2018,kucharski2020}. In particular, most extensions were proposed for adapting the SIR model to specific case studies \citep{liu2012,peng2020} or to include time-varying coefficients with the scope to estimate epidemic dynamics \citep{hooker2011,chavez2017,fang2020}.

This paper considers a flexible extension of the SEIR model, which incorporates the temporal dynamic connected to the transmission rate parameter, which is one of the most critical indicators for epidemiologists and at the basis of the basic reproduction number $R_O$. Also, this method allows us to make considerations about both the trend and the prediction of the number of infected cases  to evaluate how any possible influencing factors like as the presence of a vaccine or restriction measures taken by the central authorities can affect an epidemic outbreak \citep{haas2020}.

The proposed method is applied to the 2019–20 coronavirus pandemic, the COronaVIrus Disease 2019 (COVID-19), caused by a Severe Acute Respiratory Syndrome CoronaVirus 2 named SARS‑CoV‑2 \citep{who2020}. The World Health Organization declared the outbreak to be a Public Health Emergency of International Concern on January 30, 2020 and recognized it as a pandemic on March 11, 2020 \citep{eurosurveillance2020}.

It is worth to mention, several studies are known in the literature concerning SEIR models with different time-varying parameter specifications \citep{hong2020,petropoulos2020,piccolomini2020,wu2020,zhang2020}.
Our proposal differs in a statistical model consistent with counting data, a semiparametric (and therefore more flexible) specification of the time-varying parameters. We also pay more attention to assessing the uncertainty of estimates.

To study how country-based mitigation measures influence the course of the  SARS-CoV-2 epidemic \citep{anderson2020will}, we have looked to the on-going epidemic in the three countries (Italy, Iceland and the United States of America) where the adopted mitigation measures have been different \citep{remuzzi2020, gudbjartsson2020,dong2020}.

The reference datasets are presented in Section \ref{sec:data}, while the  proposed model and the statistical inference  are illustrated in Section \ref{sec:methods} and Section \ref{sec:statinf}. In Section \ref{sec:results}, we collect our results for the different countries  with a proposal for the forecast. We end the paper with a brief discussion in Section \ref{sec:discussion}. 
\section{COVID-19 datasets}\label{sec:data}
According to the aim of this paper, we used the screening data of daily new cases and the number of the total amount of positive cases of SARS CoV-2 according to three countries in  different phases of an epidemic outbreak. 
For each country, we considered the data in a time window that starts about 15 days before the adoption of restrictive measures and the sources of the data are described in the Appendix.

At the time of writing, the United States  of America (US)  was in the growing phase of the epidemic outbreak with an increasing trend of new cases.  In the US every single state could decide the need to adopt  stay at home measures. In those states that adopted restrictions, we registered different starting dates: the earliest state was Puerto Rico (March 15, 2020), followed by California (March 19, 2020) and New York (March 20, 2020) where the highest increase of new cases was subsequently recorded. Every single state could adopt a different and inhomogeneous panel of restrictive measures and in some States, as Arkansas, Iowa, Nebraska, and North Dakota, the local government never issued stay-at-home orders \citep{Lyu:2020}.  Data were analyzed from March 4, 2020, to April 27, 2020, for a total of 55 days of observation.
  At the end of the considered period, we reported 820,514 current positive cases, 56,259 deaths, and a total of 988,197 confirmed cases nationwide.

 Italy was a  country in the middle phase where a first stabilization of the SARS-CoV-2 incidence was reported after the restriction measure, and, at the time of writing, we observed the beginning of a decreasing trend. 
 The Italian Government adopted a national home lockdown restriction on March 9, 2020, for all the population followed by more severe measures on March 11, and ordered all nonessential businesses to close on March 22.
 Data were analysed from February 23, 2020, to April 28, 2020, for a total of 66 daily  observations.
At the end of the considered period, we reported 105,813 current positive cases,  26,977  deaths, and a total of  199,414 confirmed  cases. 

Finally, Iceland was a country in the ending phase,  where after a stabilization, the incidence of new cases was going down, and the current epidemic outbreak was probably going to the end. 
 The Iceland government adopted stricter measures to slow down the spread of SARS-CoV-2 on March 16, 2020, with an active searching strategy of new cases that lead to perform oropharyngeal swabs to about 10\% of the entire population.
 We considered data from February 29, 2020, to April 27, 2020, for a total of 59 days of observation.
At the end of the considered period, we reported 158 current positive cases, 10  deaths, and a total of 1,792  confirmed cases.

\section{SEIR model with time-varying coefficients}\label{sec:methods}
\textbf{\subsection{SEIR model}}
We start introducing the SEIR model, which is one the most used extensions of the standard SIR model, an Ordinary Differential Equation (ODE) based epidemiological model \citep{kermack1927}. 
Traditionally the SEIR model divides a population of hosts into four classes: Susceptible (S), Exposed (E), Infected (I) and Recovered (R).
However in our framework, the last class should collect all the subjects which move outside the  (I) status, i.e. recovered and deceased; for this reason, hereafter we denote  (R) as Removed status.
The model describes how the different portions of the population change over time $t$.
In the standard SEIR model, deaths are modelled as flows from the $S$, $E$, $I$, or $R$ compartment to outside, because natural deaths are normally not monitored.
If $S$, $E$, $I$, and $R$ refer to the numbers of individuals in each compartment, then these ``state variables'' change according to the following system of differential equations:

\begin{subequations}\label{eq:SEIR}  
	\begin{align}
	\frac{d}{dt}S(t) &= \mu (N-S(t))-\beta \frac{S(t) I(t) }{N}\\  
	\frac{d}{dt}E(t) &= \beta \frac{S(t) I(t)}{N}-(\mu+\sigma) E(t)    \\
	\frac{d}{dt}I(t) &=\sigma E(t)- (\mu+\gamma)I(t) \\
	\frac{d}{dt}R(t) &= \gamma\,I(t)-\mu\,R(t)
	\end{align}
\end{subequations}
In the equations \eqref{eq:SEIR} $N$ is the total population, $\mu$ is the mortality rate, $\beta$ is the transmission rate, $\sigma$ is the exposed to infectious rate, and $\gamma$ is the removal rate that can broadly assumed to be the sum of $\gamma_R +\gamma_D$, where $\gamma_R$ and $\gamma_D$ are the recovery and the mortality rate, respectively.

In general, $\beta$  is called transmission rate that is  the number of people that a positive case infects each day; in our settings, $\beta$ is defined as equal to $a b$, where $a$ is the contact rate that is the average number of contacts per person in a day while $b$ is the probability of disease transmission in a single contact. However, $a$ and $b$ cannot be identified on the basis of the current information. The ratio ${S(t)}/{N}$  permits to adjust $\beta$ taking to account people who cannot infect each other.

The parameters $\sigma$ and $\gamma$ are strictly dependent on the specific disease causing the epidemic and on the fraction of susceptible population. The parameter $\sigma$ is set equal to $\eta^{-1}$ where $\eta$ is the incubation period which may be higher for asymptomatic subjects; $\gamma$ is the recovery rate calculated as $\gamma=\rho^{-1}$ where $\rho$ is the average duration of the disease in days.

Moreover, unlike the full specification, we do not consider the effect of births in model \eqref{eq:SEIR} and, therefore,  $\sigma E (t) $ represents the number of new infected.

Based on this parametrization we can define  the reproduction number, $R_O$, as
$$R_O=\frac{\beta \sigma}{(\gamma+\mu)(\sigma+\mu)}.$$
The index conveys the strength of contagious in an epidemic outbreak. 
In the case of both $\sigma$ and $\gamma \gg \mu$, $R_O$ can be approximated by  ${\beta}/{\gamma}$.

\subsection{Time-varying parameter specification}

The standard SEIR model does that the parameters $\mu, \beta, \sigma$, and $\gamma$  are time-invariant. However, the characteristics of an epidemic suggest us that these parameters can vary. In particular, the overall mortality rate $\mu$ may increase if the number of deaths in a population directly or indirectly attributable to the disease (i.e., the insufficient capacity of health services) rises. The $\beta$ rate may also vary according to social distancing policies or, even, the isolation of infected people.

We aim to  evaluate as the actions taken by the governments, and how a different degree of travel restrictions, social distancing or limitation of the people movement can affect the epidemic curve of the infectious population. Our working hypothesis is that if there is an effect of the actions, they only affect  the transmission rate of the epidemic, $\beta$. For this reason, we propose to modify this parameter over time, namely    
\begin{subequations}\label{eq:SEIRvar}  
	\begin{align}
	\frac{d}{dt}S(t) &= \mu (N-S(t))-\beta(t) \frac{S(t) I(t) }{N}\\  
	\frac{d}{dt}E(t) &= \beta(t) \frac{S(t) I(t)}{N}-(\mu+\sigma) E(t)    \\
	\frac{d}{dt}I(t) &=\sigma E(t)- (\mu+\gamma)I(t) \\
	\frac{d}{dt}R(t) &= \gamma\,I(t)-\mu\,R(t)
	\end{align}
\end{subequations}
Since the function $\beta(t)$ takes positive values, in the estimation step we consider the following log-linear specification
\begin{equation}\label{eq:splines}
\log(\beta(t))=\sum_{k=1}^K \psi_{k} N_k(t),
\end{equation}
where $N_k(t)$, $k=1,\ldots,K$,  are $K$ natural cubic spline basis functions evaluated at $K-2$ equally spaced knots in addition to the boundary knots.   
 
The representation in \eqref{eq:splines} has the advantage that the estimation of $\beta(t)$ reduces to the estimation of the coefficients  $\psi_k$.
 We refer to the next subsection for a short discussion about the number  of knots and their positions.
The time-dependent transmission rate $\beta(t)$ allows us to define a time dependent version $R_O$ (the basic reproduction number) as follows
$$R_O(t)=\frac{\beta (t) }{\gamma}.$$
This index permits to evaluate the strength of contagious over a temporal window comparing $\beta (t)$ with the removal rate $\gamma$. In order to constraint $\gamma$ between 0 and 1, in the estimation process we reparametrize $\gamma$ as $\gamma$=$\frac{exp(\gamma^*)}{1+exp(\gamma^*)}$.

The  system \eqref{eq:SEIRvar} is a system of nonlinear ODEs, which must be solved numerically. In this paper we use the ODE solver lsode  \citep{Hindmarsh:1983} as it has implemented in the R package \texttt{deSolve}. 
If we suppose that $\mu$ and $\sigma$ are known parameters, the (numerical) solutions $S(t;\theta)$, $E(t;\theta)$, $I(t;\theta)$, an  $R(t;\theta)$ depend on the (vector of) parameters $\theta=(\psi_1,\ldots,\psi_K,\gamma^*)$.

\section{Statistical inference}\label{sec:statinf}
Different agencies in the world that daily update and publish datasets of epidemic data that contain at least three time series: the total number of infected, the number of dead, the number of recovered (see section \ref{sec:data} for more details). We derive from these time series   the  daily number of current positive cases $Y(t)$ and the daily number of new positive cases $Z(t)$, recorded at day $t$, $t=1,\ldots,T$. 

Usually, the time series are supposed to be realizations of 
a stochastic version of the compartmental models. 
The different versions can  be broadly classified into continuous models and discrete models. In the first group
fall the continuous-time Markov chains (CTMCs) and the stochastic differential equations (SDEs) \citep{allen:2008}.
In the second group a discrete-time approximation to the stochastic continuous-time model is considered \citep{Lekone:Finkenstadt}.
  There exists an extensive literature on 
calibrating the  stochastic models against time-series with different inferential approaches \citep{Finkesdadt:Grenfell:2000,Ionides:2006,hooker2011,andersson2012stochastic,Dureau:2013}.

Instead in this paper we follow the simplest idea  
 that the solutions of the system \eqref{eq:SEIRvar} are actually the expectations at days $t=1,\ldots,T$ of as many counting random variables. More precisely we model the observed counts $\{Y(t),Z(t)\}$,  as
\begin{subequations}\label{eq:poisson}
	\begin{align}
	Y(t) \sim &\operatorname{Poisson} (I(t;\theta))\\
	Z(t) \sim &\operatorname{Poisson}(\sigma E(t;\theta)).\qquad 
	\end{align}
\end{subequations}
Then the estimate of the parameter $\theta$ already defined are obtained by maximizing the independence log-likelihood \citep{Chandler:Bate:2007} 

\begin{eqnarray}\label{eq:cl}
	cl(\theta) &=& \sum_{t=1}^T Y(t)\log I(t;\theta) -I(t;\theta)+ 
	Z(t)\log  (\sigma E(t;\theta)) -\sigma E(t;\theta)\\
	&=&\sum_{t=1}^T cl(\theta;t).\nonumber 
\end{eqnarray}

Note that  $CL(\theta)$ is not a `true' log-likelihood but an instance of a composite likelihood \citep{Lindsay:1988} since it does not seem reasonable to assume that $Y(t)$ and $Z(t)$ are mutually and temporally independent.
However, even though the model is not correctly specified, the maximum composite likelihood estimator, $\widehat{\theta}$, is still a consistent and asymptotically Gaussian estimator with  asymptotic variance $V({\theta})$ under mild conditions \citep{Chandler:Bate:2007,Jacod:Sorensen:2018}.

The variance $V(\theta)$  can be  estimated by the sandwich estimator
$
\hat{V}=\widehat{B}^{-1} \widehat{M}\widehat{B}^{-1'}
$. The `bread' matrix is given by
$\widehat{B}={T^{-1}}\sum_{t=1}^T\nabla u(\hat\theta;t)$ with 
$u(\hat\theta;t)=\nabla cl(\hat\theta;t)$.
In the presence of time-dependence  the `meat' matrix $\hat{M}$ is given
by the  heteroskedasticity and autocorrelation consistent (HAC) estimator
$$
\hat{M}=T^{-1}\sum_{t=1}^T\sum_{s=1}^Tw_{|t-s|}\nabla u(\hat\theta;t)\nabla u(\hat\theta;t)^\top
$$ where $w = (w_0 , \ldots , w_{T-1})$ is a vector of weights \citep{Andrews:1991}.

With the aim of forecasting the spread of the epidemic   outside the observed period,  the number and the positions of knots in \eqref{eq:splines} play a crucial role. 
The higher the number of nodes, the less smooth the function $\beta(t)$. In this way, however, there is a risk of over-fitting the data.
On the other hand, the trend of the $\beta(t)$ outside the observation time interval is mainly determined by the basis functions corresponding to the boundary knots.

We select the number of knots  by maximizing the Composite Likelihood Information Criterion (CLIC) \citep{varin2005}
$$
CLIC(\widehat{\theta})=cl(\hat\theta)+\operatorname{tr}( \widehat{B}^{-1} \widehat{M}).
$$ 
The criterion  has a strong analogy with the Akaike Information Criterion (AIC). In fact $cl(\hat\theta)$ measures the goodness-of-fit similarly to the log-likelihood and the penalty tr($\widehat{B}^{-1} \widehat{M}$) reduces to $-(K+1)$ if the  model \eqref{eq:poisson} is correctly specified, i.e. if  equation \eqref{eq:cl} is the `true' log-likelihood.

We could locate the internal knots to reflect policy interventions.
 However, it is very difficult to hypothesize the immediate effects of these policies and a simpler choice has been to place temporally equally spaced nodes.
  As for the boundary knots, it was chosen to place them at the beginning of the period and one week after the last observation available to obtain more stable estimates in the forecast period.

\section{Results}\label{sec:results}
\begin{figure}[!ht]
	\begin{center}
		\includegraphics[width=\linewidth]{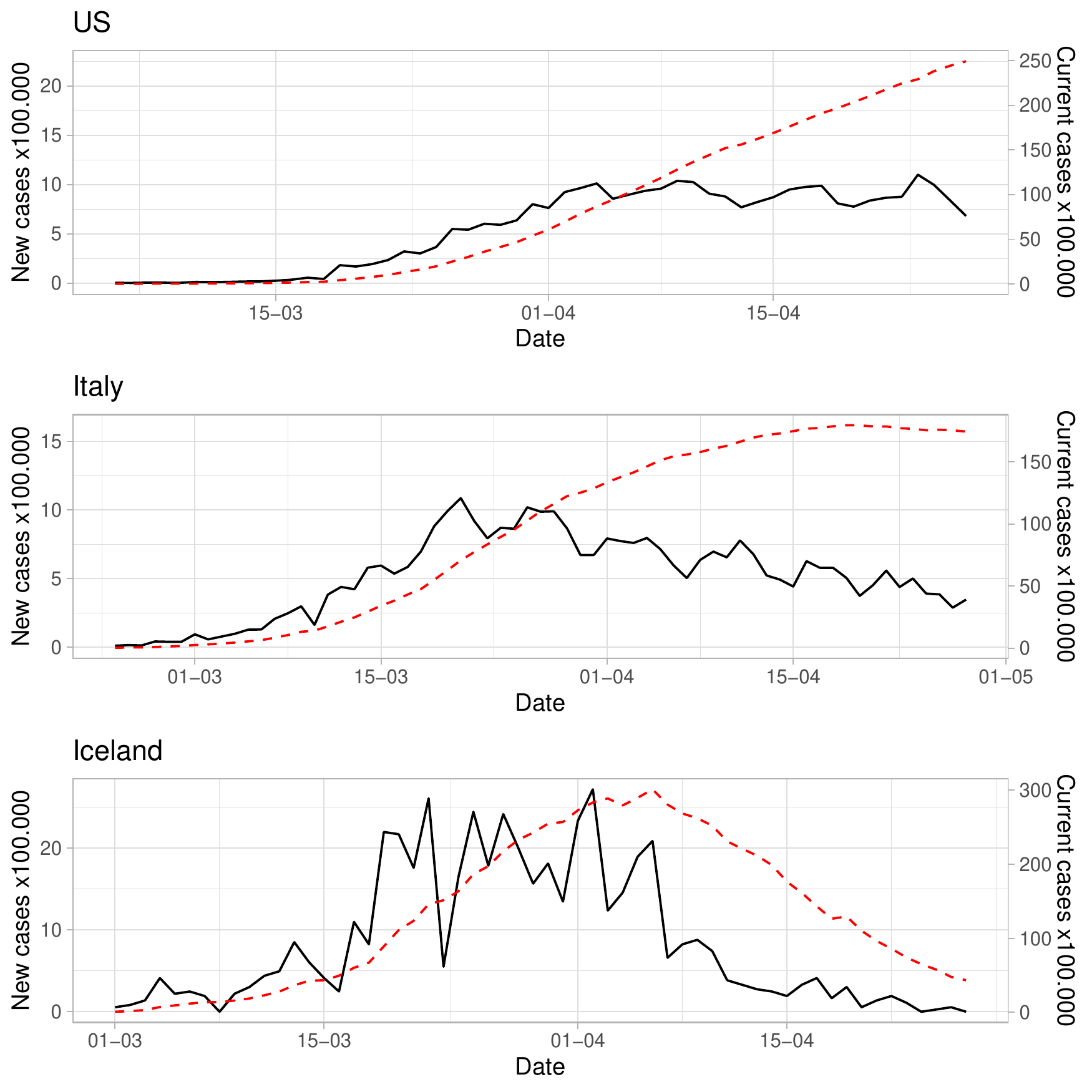} 				
	\end{center}
	\caption{Daily new  (solid black line) and current (dashed red line) cases  of SARS-CoV-2 in (a) the US, (b) Italy and (c) Iceland.}\label{fig:Figure1}
\end{figure}

The epidemic outbreak showed different patterns in the selected time window:  the reported SARS-CoV-2 cases in the US were rapidly increasing, reaching a peak and subsequent  stabilization of the number of daily new cases with an incidence of about 10 cases x100.000 inhabitants; in Italy, a drop up to 2.5 daily new SARS-CoV-2 cases x100.000 was reported, after an initial growth which reached a peak of incidence of about 10 cases x100.000 people similar to the US; in Iceland, the incidence of new SARS-CoV-2 cases knew a huge peak ($\approx$ 25 cases x100.000 people), then a decreasing trend and finally a limited number of new cases in the last considered day (Figure \ref{fig:Figure1}). A the end of the temporal window prevalence of the disease was quite different among the three considered countries: we registered 250, 180, and 45 current SARS-CoV-2 cases x100.000 people in the US, Italy, and Iceland, respectively.

In the literature  the incubation duration  of the SARS-CoV-2 was estimated as $ { \eta}={5.2}  $ \citep{wang2020} and therefore we set the specific parameter $\sigma=1/\eta=0.192$.  

The overall mortality rate $\mu$ was calculated  as  ${1}/{(\mbox{lifespan})}={1}/(365.25 \times\mbox{LE})$ where the Life Expectancy (LE) is 78.5 years in the US, 83.2 years in Italy  and 82.2 years in Iceland, respectively.
The total population (N) in 2020 was is 329.23 (US), 60.32 (Italy) and 0.36 (Iceland) million of inhabitants.
The starting values $S(0)$, $E(0)$, $I(0)$ and $R(0)$  for the numerical resolution of the system \eqref{eq:SEIRvar} were set as follows
\begin{itemize}
	\item $I(0)=Y(1)$ i.e. the number of currently infected on the first day of the dataset (US: 142,  Italy: 155, Iceland: 1);
	\item $R(0)$  equal to the number of currently recovered on the first day of the dataset (US: 7,  Italy: 0, Iceland: 0);
	\item $E(0)= Z(1)/{\sigma}$ where $Z(1)$ is the number of new infected on the first day of the dataset (US: 68,  Italy: 66, Iceland: 2);
	\item $S(0)=N-E(0)-I(0)-R(0)$.
\end{itemize} 
%
%

We have tried several values for the number of basis function $K$, i.e. from 3 to 8, and we found that $K=5$ and $K=3$ minimize the value of CLIC for Italy/US and Iceland, respectively.

\begin{figure}[!ht]
	\begin{center}
		\includegraphics[width=\linewidth]{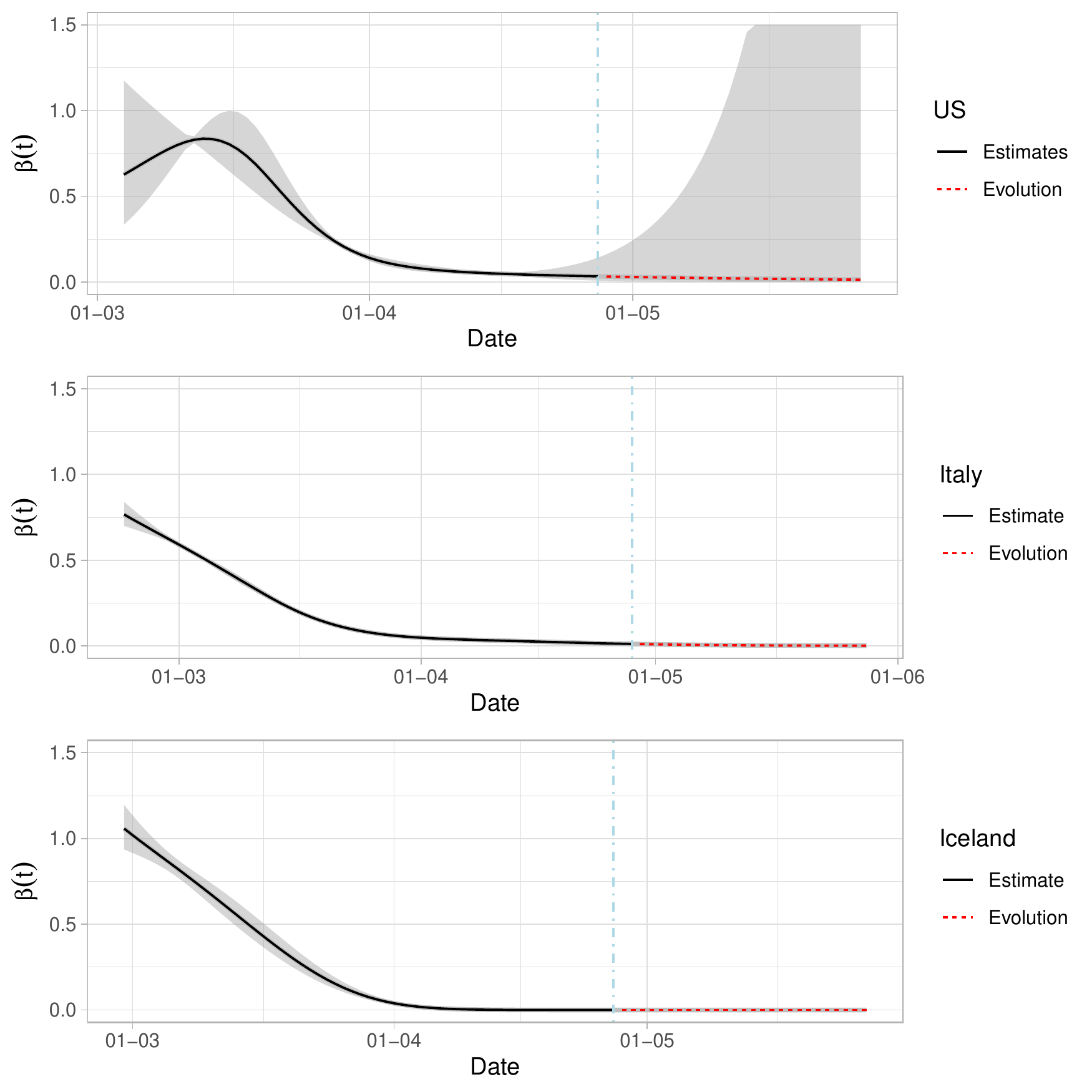} 				
	\end{center}
	\caption{The estimate of  $\beta(t)$ curves and 95\%  confidence bands (in grey colour) for the US, Italy and Iceland. The dashed line represents the 30-day predicted evolution.}\label{fig:Figure2}
\end{figure}

The estimate of  $\beta(t)$  (see Figure $\ref{fig:Figure2}$) showed an overall decreasing pattern of the transmission rate across the selected countries. In particular, in the US the estimate of ${\beta(t)}$ reached a peak close to $0.8$ after about ten days by the beginning of the epidemic outbreak, denoting an uncontrolled situation, moving to values of approximately $0.1$ after about $45$ days of the epidemic, with a predicted scenario of a slightly decreasing trend and a great amount of uncertainty.
In Italy, the estimate of ${\beta(t)}$ was moving from initial values of $0.75$ to value close to $0.05$ after about 50 days of observations. The estimate of $\beta(t)$  were lower than those reported for the US. The estimate of  $\beta(t)$ in Iceland showed a fast decreasing trend from values a bit over $1.0$ to about 0 at day 40. The 30-days prediction for $\beta(t)$ is pratically zero, denoting the end of the  current phase of the epidemic. 

\begin{table}
	\centering
	\begin{tabular}{lcc}
		\hline
		Country &   $\gamma$ &  (95\% CI)  \\ 
		\hline
		US & 0.012 & [0.009-0.015]\\
		Italy  & 0.025 & [0.023-0.027]\\
		Iceland & 0.080 & [0.063-0.101]  \\
		\hline
	\end{tabular}
	\caption{Estimated values  of the $\gamma$ parameter and its relative 95\% Confidence Interval.}
	\label{tab:table1}
\end{table}

The estimates  of $\gamma$ ranged from a low rate in the US and Italy, $0.012$ and $0.025$ respectively,  to a higher rate in  Iceland, $0.080$. Then the removal duration, i.e. the reciprocal of $\gamma$, was estimated at 85.0 days (95\% CI:  65.5-110.5) for the US, at 40.2 days  (95\% CI: 37.6-43.1) for Italy and at 12.5 days  (95\% CI: 9.9-15.9) for Iceland.

\begin{figure}[!ht]
	\begin{center}
	\includegraphics[width=\linewidth]{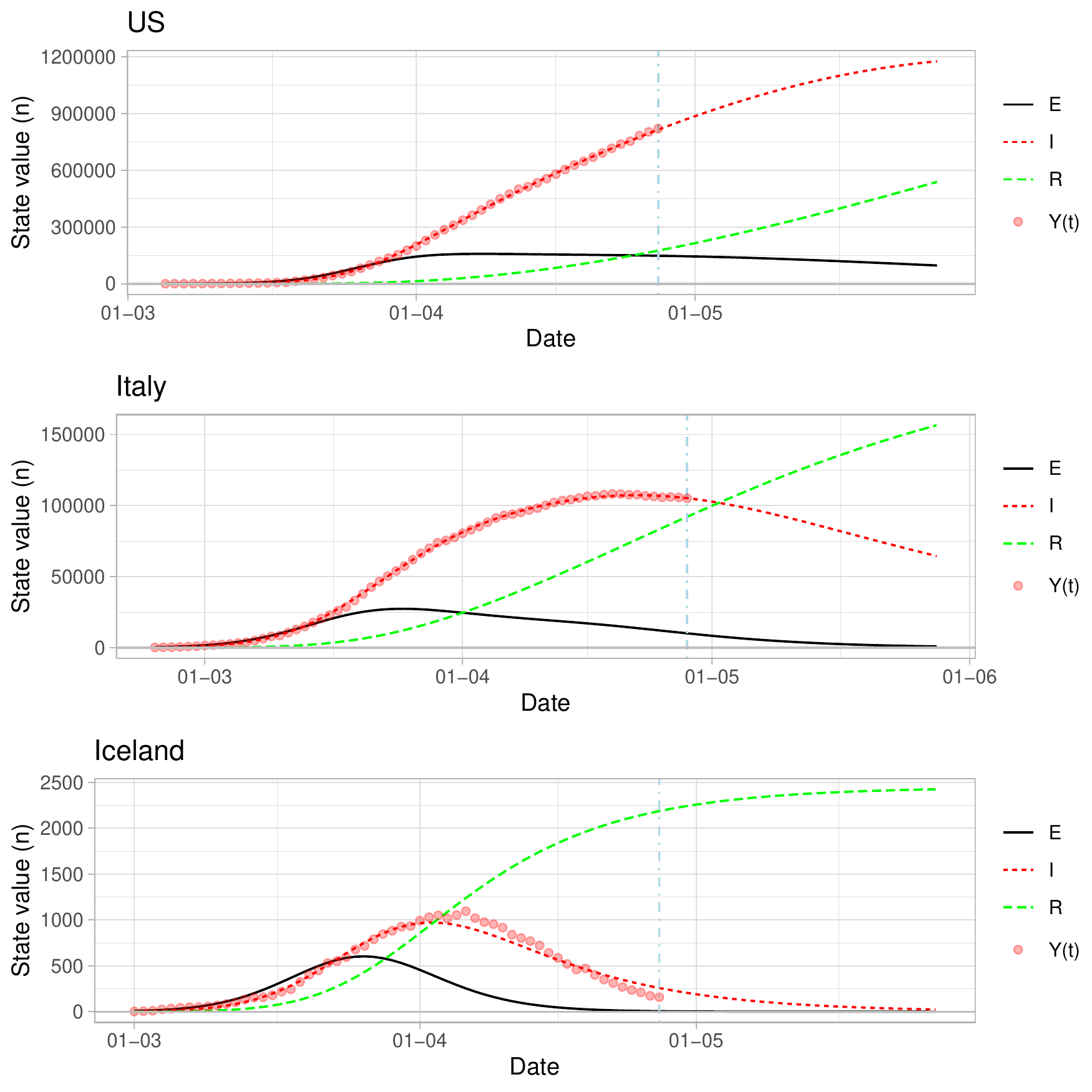} 			
		
	\end{center}
	\caption{The expected number of Exposed, Infected and Recovered subjects for the US, Italy and Iceland, based on the model parameter estimates. The dotted points indicate the observed number of infected cases. 
	}\label{fig:Figure3}
\end{figure}

\begin{figure}[!ht]
	\begin{center}
			\includegraphics[width=\linewidth]{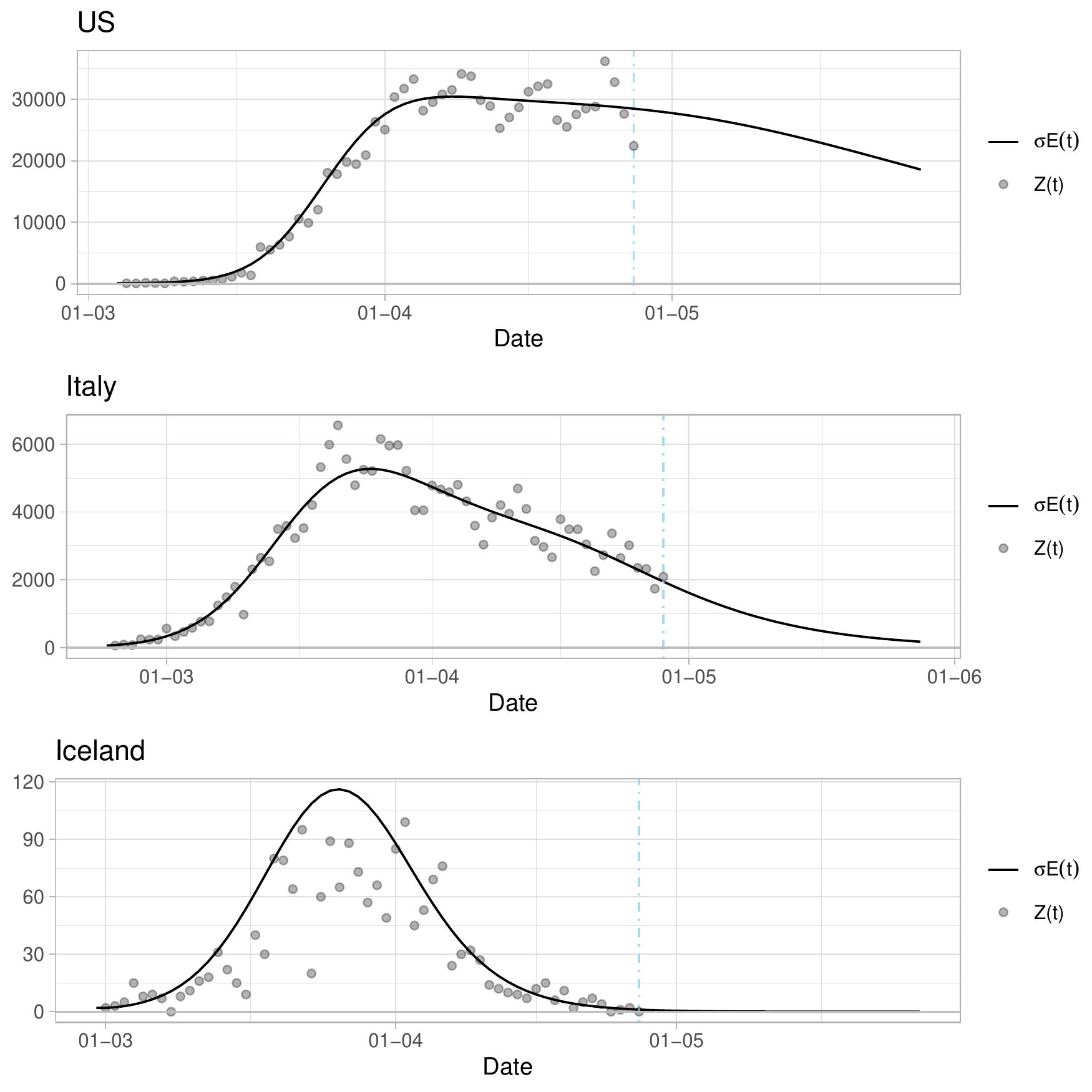} 
	\end{center}
	\caption{The expected number of new infected cases  for the US, Italy and Iceland, based on the model parameter estimates. The dotted points indicate the observed number of new infected cases.
	The dashed line indicates the last day observed.
}\label{fig:Figure4}
\end{figure}

The model fitting was deemed satisfactory (Figure $\ref{fig:Figure3}$ and Figure $\ref{fig:Figure4}$) both with respect to the number of new cases and to the cumulative positive cases.  
Our major findings were:  in the US, the current positive cases were going to increase, reaching a probable maximum after the window of the next 30 days. In Italy, the epidemic outbreak had known its maximum in the number of positive cases around April 20th, and the tendency was for a slight decline. In Iceland, the peak of positive cases was registered on April 10th, associated with a rapid decreasing phase and a low number of new cases in the last observed days; in this case, the SARS-CoV-2 epidemic was going to be overcome approximately at the end of May.

\begin{figure}[!ht]
	\begin{center}
		\begin{tabular}{c}
			\includegraphics[width=\linewidth]{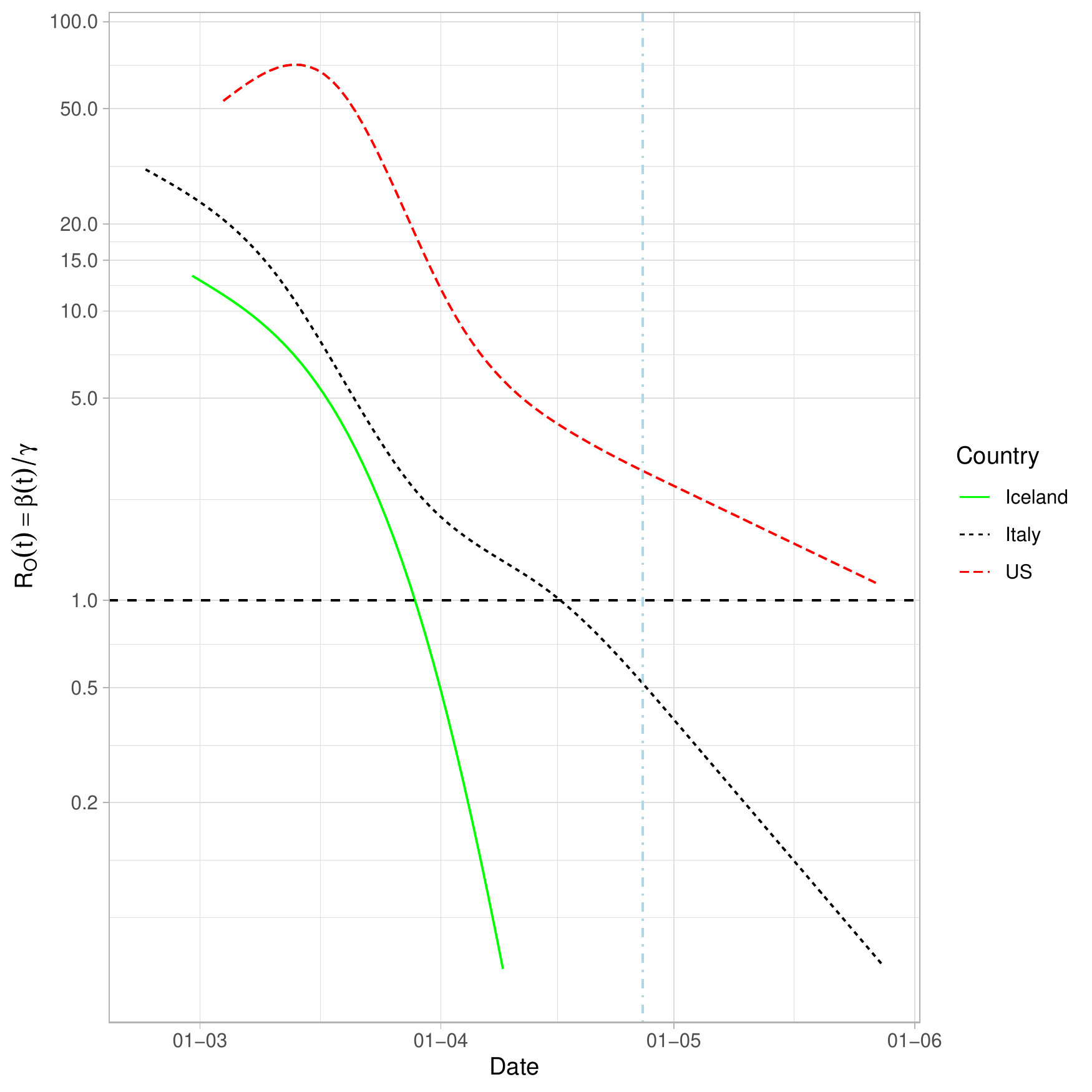} 
		\end{tabular}
	\end{center}
	\caption{Estimated $R_O(t)$  values for the US, Italy, and Iceland and predicted evolution. The Y-axis is in the log-scale. 
		The dashed lines indicate $R_O=1$ and the last day observed, respectively.}\label{fig:Figure5}
\end{figure}

The estimated trend for $R_O(t)$ appears quite different among the selected countries (see Figure \ref{fig:Figure5}). The value $R_O(t)=1$ was reached on different dates in Iceland (March 28th) and in Italy (April 16th), while in the US is expected to be achieved only in the end of May.

\section{Discussion}\label{sec:discussion}
International health organizations recommend to implement public health and social measures to slow or stop the spread of SARS-CoV-2, reaching the full engagement of all members of society \citep{who2020}.
Countries have adopted different public health and social measures depending on the local specific historical evolution of the SARS-CoV-2 pandemic and on their health system capacity. Our analysis considers data of three countries, the US, Italy and Iceland which have, on one side, different geographic and demographic characteristics and, on the other one, as many dissimilar approaches in terms of public health policies and restrictive measures concerning the on-going epidemic.

Our proposal allows us to estimate an epidemiological SEIR model with a time-varying transmission rate ($\beta(t)$) with the scope to assess the timeline and the strength of the effects produced by the adopted restrictive measures. The removal rate $\gamma$ was estimated, considering both two different time series (daily new and current positive cases). We avoided considering estimates of clinical SARS-CoV-2 recovery rate and specific mortality rate calculated by others, with the scope to comment on the knowledge provided by the analysed data on the removal rate estimated by our model.

In the US, the transmission rate of SARS-CoV-2 was very high at the beginning of the considered temporal window, and its reduction appears to be later and slower in comparison with those observed in Italy and even more in Iceland. This difference may be viewed as a result of the approach adopted by the US in the epidemic onset based only on a limited and non-homogeneous containment measures \citep{parodi2020}.

The adoption of this  strategy in the US was a decision of particular  importance since the COVID-19 onset began a few days later than Italy, where, on March 9, 2020, the Italian Government set Europe's first nationwide restriction on movement due to the incoming SARS-CoV-2 epidemic. The estimates on the Italian transmission rate confirm the control of the epidemic wave after approximately 20 days of home restrictions, but with a high mortality toll in comparison with the preceding Chinese epidemic \citep{rubino2020}. A social distancing and passive testing of symptomatic cases was the Italian strategy to contain the epidemic. Positive cases with few symptoms were confined in home isolation. However, there was a consistent amount of asymptomatic, which remained undetected,  contributing to	 spread the epidemic \citep{lavezzo2020, flaxman2020}.

Iceland has the advantage that the epidemic outbreak started later than Italy; we observed that the Icelandic transmission rate quickly moved to values close to 0 after only 15 days of the restrictive measures \citep{gudbjartsson2020}. These results were mainly attributable to an active searching strategy of asymptomatic positive cases organized by the national health service, which lead to be tested about 6\% of the Iceland population at the date of April 2, 2020. However, the presence of a free voluntary private screening program estimated that the fraction of undetected infections by the Icelandic health service ranged from 88.7\% to 93.6\% \citep{stock2020}.

The comparison between the US, Italy, and Iceland was certainly affected by a regional/state variation to COVID-19 response. However this administrative-level granularity plays a role in the diffuseness of non-pharmacological health measures in the first phase of an epidemic outbreak. A faster and more aggressive response of each local administrative unit helps to contain the contagious spread and to achieve a relative control of the disease \citep{Lancet:2020}. Despite fears of the negative consequences on their economy, Italy and Iceland experienced a contagion control in a relative short period. With a rapid and structured stay-at-home order and an assertive infection control measures Iceland reduced the required time to flatten the curve.

The estimated values for $\gamma$ reflects both the locally adopted swab policy and the specific phase of the epidemic wave: in fact, the active monitoring in Iceland provide a reliable value for the removal rate that is of about 12 days in line with that measured in China (14 days, \citep{wang2020}). In Italy, the controlling strategy implies that after a first positive swab test, a control swab will be repeated after a period of home isolation and this fact implies a longer time to obtain the healing confirmation.

In the US the low number of removed subject in comparison with the high increase in the incidence makes challenging a reliable estimation of the removal rate.

The proposed model has become a standard approach to estimate the transmission rate in a dynamic context \citep{hong2020,petropoulos2020,piccolomini2020,wu2020,zhang2020,Godio:2020}. The model has the double scope of having a real-time monitoring and of supplying possible evolution scenarios. The estimated fluctuations of $\beta(t)$ were driven by gradual changes in the behaviour of the population at risk as a consequence of the adopted restrictions. Respect to other specifications, our approach has the advantage of employing a basis of splines that allow us a high grade of flexibility for the estimation. 

We estimated the parameters for $\beta (t)$ and $\gamma$ through a composite likelihood considering the information provided by both the occurrence of new SARS-CoV-2 cases and  the current positive cases; in order to cope with the possible presence of heteroscedasticity and autocorrelation in the data, we estimated consistent standard errors combining a sandwich variance estimator and a HAC correction.
Even though the model formulation has some ancestors, our proposal differs in two aspects from the aforementioned literature: we allow the data to indicate the shape of the function $\beta(t)$ using a semiparametric approach. This feature could help to identify the best health intervention policy in a country; HAC variance estimators permits to reduce the bias allows to correct the underestimation of the variance of the estimator and therefore to produce future scenarios with a more appropriate margin of uncertainty.

There are several other limitations to our analysis. We used plausible biological SARS-CoV-2 parameters for the SEIR model based on updated numbers (i.e., $\sigma$), but these values may be refined as more comprehensive data become available. The predicted values for $\beta(t)$  are valid only in the absence of future changes to the restrictions, which is not likely to happen  if an intermittent social distancing measures will be adopted \citep{ferguson2020}. 

Our results point out that the transmission rate in the US, Italy, and Iceland showed a decline after the introduction of restriction measures. Despite this common trend, some differences in terms of timeline and impact are present. In particular, US experts argue that more helpful tools are needed in order to reach the control of the epidemic wave \citep{parmet2020}.

The adoption of restrictive measures results in flatten epidemic curves and thus the distribution of the SARS-CoV-2 cases in a more extended period, respect to an uncontrolled epidemic outbreak. In the absence of a specific vaccine, the high number of susceptibles and the relaxation of restrictions taken represent a cause of future outbreaks.
\newpage
\section*{Appendix}

\subsection*{Data sources accessed on April 28, 2020.}
\begin{description}
		\item[Iceland:] John Hopkins University (\texttt{github.com/datasets/covid-19}), \citet{csse2020}.
		\item[Italy:] Italian Civil Protection, (\texttt{github.com/pcm-dpc/COVID-19}),  \citet{morettini2020}.
	\item[US:] John Hopkins University (\texttt{github.com/datasets/covid-19}), \citet{csse2020}.

\end{description}	
\subsection*{Software}
The statistical analysis  was carried  using the R software \citep{R2019} and some its   package: the minimization of the previous quantity was performed by means of a non-linear minimization process using the function \texttt{nlm}; package \texttt{deSolve} to resolve the standard ODE and package \texttt{ggplot2} to enhance the quality of  the figures. Results and figures  can be reproduced using the companion code in \texttt{github.com/Paolin83/SARS-CoV-2\_SEIR\_TV\_model}.

\bibliographystyle{apacite}

\bibliography{mybib.bib}
\end{document}